# High Entropy Oxide Relaxor Ferroelectrics

*Yogesh Sharma[#,†,\*], Min-Cheol Lee[†], Krishna C. Pitike[#], Karuna K. Mishra[¶], Qiang Zheng[#], Xiang Gao[#], Brianna L. Musico[‡,§], Alessandro R. Mazza[#], Ram S. Katiyar[¶], Veerle Keppens[§], Matthew Brahlek[#], Dmitry A. Yarotski[†], Rohit P. Prasankumar[†], Aiping Chen[†], Valentino R. Cooper[#], and T. Zac Ward[#]*

[#]*Materials Science and Technology Division, Oak Ridge National Laboratory, Oak Ridge, TN 37831, USA*

[†]*Center for Integrated Nanotechnologies (CINT), Los Alamos National Laboratory, Los Alamos, NM 87545, USA*

[¶]*Department of Physics and Institute for Functional Nanomaterials, P.O. Box 70377, University of Puerto Rico, San Juan, PR 00936-8377, USA*

[‡]*Sigma Division, Los Alamos National Laboratory, Los Alamos, NM 87545, USA*

[§]*Department of Materials Science and Engineering, University of Tennessee, Knoxville, TN 37996, USA*

[\*]E-mail: ysharma@lanl.gov

## Abstract

Relaxor ferroelectrics are important in technological applications due to a strong electromechanical response, energy storage capacity, electrocaloric effect, and pyroelectric energy conversion properties. Current efforts to discover and design new materials in this class generally rely on substitutional doping of known ferroelectrics, as slight changes to local compositional order can significantly affect the Curie temperature, morphotropic phase boundary, and electromechanical responses. In this work, we demonstrate that moving to the strong limit of compositional complexity in an $AB$O$_3$ perovskite allows stabilization of novel relaxor responses that do not rely on a single narrow phase transition region. Entropy-assisted synthesis approaches are used to create single crystal Ba(Ti$_{0.2}$Sn$_{0.2}$Zr$_{0.2}$Hf$_{0.2}$Nb$_{0.2}$)O$_3$ [Ba(5$B$)O] films. The high levels of configurational disorder present in this system is found to influence dielectric relaxation, phase transitions, nano-polar domain formation, and Curie temperature. Temperature-dependent dielectric, Raman spectroscopy and second-harmonic generation measurements reveal multiple phase transitions, a high Curie temperature of 570 K, and the relaxor ferroelectric nature of Ba(5$B$)O films. The first principles theory calculations are used to predict possible combinations of cations to quantify the relative feasibility of formation of highly disordered single-phase perovskite systems. The ability to stabilize single-phase perovskites with such a large number of different cations on the *B*-sites offers new possibilities for designing high-performance materials for piezoelectric, pyroelectric and tunable dielectric applications.

# 1. INTRODUCTION

The effect of compositional disorder on order parameter coupling in $ABO_3$ perovskites is broadly important to a range of fundamental and applied interests.[1–4] Disorder in these oxides has been realized as a key concept to understand the underlying physics of exotic phenomena emerging from complex interactions of order parameters—examples include, quantum critical behavior in superconductors,[5] controlling magnetoresistance in double perovskites,[6] and stabilizing polar nanoregions in ferroelectrics.[7]

The influence of local compositional or cation disorder on mesoscopic properties is not well studied, but the existing examples are promising. As in the example of the relaxor ferroelectrics (RFEs), where modifying the type and magnitude of disorder can significantly affect the Curie temperature, morphotropic phase boundary (MPB), polar nanodomains, and electromechanical properties.[7–13] RFEs show disordered polarization, where the small ordered 'polar nanoregions' are individually polarize below a temperature known as the Burns temperature.[12,14] RFEs are characterized by a broad maximum in the temperature-dependent dielectric permittivity, where a strong frequency dispersion in the dielectric response below the temperature of maximum permittivity can be observed.[7,8] RFEs are well known for their strong electromechanical response,[15] energy storage capacity,[16,17] electrocaloric effect,[18] and pyroelectric energy conversion properties.[19] The best-known member of the RFE family is the disordered $ABO_3$ perovskite crystal. In particular, chemical heterogeneities or *B*-site randomness in RFEs is commonly accepted to be intrinsic to the appearance of the relaxor state and have a notable impact on the macroscopic properties.[20] The manipulation of local cation disorder in RFEs is a known route to access hidden phases, structural phase transitions, and quantum critical point phase spaces for both fundamental studies of emergent properties and a wide range of applications.[21,22]

To create high-performance RFE perovskite oxides, chemical modification—as an extrinsic substitutional defect—has been extensively used to introduce compositional disorder and to manipulate relaxor behavior.[23–26] Chemical modifications that include multiple dopants and mixing of different perovskites in a solid-solution[16,27] are the most effective approaches to design high-performance relaxors. However, the existence of phase segregation and the formation of undesired impurity phases are significant disadvantages associated with a large number of chemical dopants and the mixing of multiple perovskites.[28–39] With the complexity of the single phase formation in multi-cation doped-RFE crystals, synthesizing single crystal RFEs with a large number of different cations on the *A*- and/or *B*-sites without disrupting crystallinity or structural phase would be an exceptional tool towards designing high-performance RFEs for piezoelectric, pyroelectric, and tunable dielectric applications.[40]

The development of an entropy-stabilization approach for synthesizing single crystal $ABO_3$ perovskite films with 5 or more cations at the *A*- and/or *B*-sites[41–45] is a promising route to explore how extreme levels of microscopic disorder perturb the polar ordering. These new high entropy oxide materials conform to standard tolerance factor considerations, where multicomponent *A*- or *B*-sites can be simply treated as the average ionic radii value.[46,47] The selection of different cations is centered on applying high configurational disorder where the random distribution of constituent elements into the cation sublattice(s) enhances the configurational entropy in such oxide solutions.[48–52] For relaxors, this approach can be especially beneficial as the role of different cations (size, electronegativity) and their random cationic distribution in the lattice can hugely impact the arrangement of polar order/disorder on multiple length scale.[8,13,53]

In this paper, our aim is to characterize the dielectric and ferroelectric properties of single-crystal high entropy perovskite oxide epitaxial Ba(Ti$_{0.2}$Sn$_{0.2}$Zr$_{0.2}$Hf$_{0.2}$Nb$_{0.2}$)O$_3$ [Ba(5$B$)O] films to understand how the strong limit of compositional complexity can be used to create new candidate relaxor dielectric and ferroelectric materials with tunable properties. The selection of the $B$-site sublattice is motivated by the general trend that changes to size and charge disorder, as well as the end-members with ferroelectric-and paraelectric-like phases can manifest complex ferroelectric relaxor behavior in Ba(5$B$)O. We discuss how the local configurational disorder influences the phase transitions, dielectric properties, mesoscopic domain structures, and Curie temperature in this highly disorder relaxor perovskite oxide. Using first principles theory calculations, we propose other possible combinations of cations to design new RFEs and quantify the relative feasibility of formation of these highly disordered single-phase perovskites.

## 2. EXPERIMENTAL SECTION

**2.1 Epitaxial film growth.** A ceramic stoichiometric target of Ba(5$B$)O was synthesized using the conventional solid-state reaction method. Pulsed laser epitaxy was used to grow Ba(5$B$)O thin films on the SrTiO$_3$ (STO) (001) and MgO (001) substrates. A KrF excimer laser ($\lambda$ = 248 nm) operating at 5 Hz was used for target ablation. Deposition optimization was performed and the optimal growth conditions were found to occur with an oxygen partial pressure of 150 mTorr at a substrate temperature of 1023 K. Same growth conditions were used to grow 7 nm conducting SrRuO$_3$ (SRO) layer on STO, before growing Ba(5$B$)O layer on SRO/STO. The detailed growth conditions were described elsewhere.[46]

**2.2 XRD measurements.** The crystal structure and growth orientation of the films were characterized by X-ray diffraction (XRD) using a four-circle high resolution X-ray diffractometer (X'Pert Pro, Panalytical) (Cu K$\alpha_1$ radiation). X-ray reflectivity (XRR) measurements were carried out to measure the thickness of the films.

**2.3 HRTEM Analysis.** Cross-sectional specimens oriented along the [100]STO direction for scanning transmission electron microscopic (STEM) analysis were prepared using ion milling after mechanical thinning and precision polishing. High-angle annular dark-field (HAADF) analysis was carried out in a Nion UltraSTEM 200 operated at 200 kV. An inner detector angle of about 78 mrad was used for HAADF observation. For determining the *B*-site cation projected displacement vectors, noise in the obtained HAADF-STEM images was reduced the average background subtraction filter (ABSF).[54] The projected atom positions were determined by fitting them as 2D Gaussian peaks by using the StatSTEM software,[55] thus the projected *B*-site cation displacement (δ*B*) was deduced as a vector between each *B*-site cation and the center of mass of its four nearest *A*-site neighbor cations.

**2.4 Dielectric and ferroelectric measurements.** For electric measurements, top circular platinum (Pt) contacts (diameter ~ 250 μm) were deposited onto the Ba(5*B*)O layers by using radio-frequency sputtering through a shadow mask. To characterize the FE properties, room temperature polarization–voltage hysteresis loops and switching current–voltage loops were recorded by using an aixACCt Thin Film Analyzer 2000. Temperature dependent dielectric permittivity and loss tangent were measured at several frequencies (100 Hz to 0.1 MHz) in the temperature range of 100-670 K with an oscillating voltage of 0.5 V, using an impedance analyzer (HP 4294A) interfaced with a programmable Joule Thompson thermal stage system (MMR technology, model# K-20, temperature stability of ± 1 K).

**2.5 Second harmonic generation (SHG) experimental set up.** Far-field reflection SHG measurements were performed with an 800 nm fundamental laser beam generated from a 250 kHz femtosecond regenerative amplifier (RegA 9000, Coherent Inc.), producing 100 femtosecond (fs) pulses. The 800 nm laser beam is incident onto the film sample at a 45°angle of incidence. We measured the reflected SHG signal from the sample for output s- and p-polarizations while tuning the polarization angle of the input fundamental 800 nm beam.

**2.6 DFT calculations.** The DFT calculations were carried out using the plane-wave-based Vienna Ab initio Simulation Package VASP[56,57] version 5.4.4, within the local density approximation (LDA) exchange−correlation functional.[58] The energy cutoff for the plane-wave basis set was 700 eV, employing projected augmented wave potentials.[59,60] An $8 \times 8 \times 8$ $k$ − point mesh was utilized for sampling the Brillouin zone for a five-atom perovskite unit cell and scaled linearly with the number of formula units present in the supercell. We have used a $2 \times 2 \times 2$ supercells for calculating mixing enthalpies of one and two component perovskites. The bulk geometry was optimized with a force convergence criterion of 1 meV/Å, and the individual components of the stress tensor were converged to $\leq 0.1$ kB. Epitaxial strain in a thin film constrained on a cubic (001)-oriented $SrTiO_3$ substrate was simulated by fixing the in-plane lattice constant $a$ of the thin film unit cell to the optimized lattice parameter of $SrTiO_3$ and allowing the out-of-plane lattice constant $c$ to relax by converging the normal stress in the out-of-plane direction to $\leq 0.1$ kB.

## 3. RESULTS AND DISCUSSION

Ba(5$B$)O films of 90 nm thickness were deposited on 7 nm conducting SRO layered (001) STO substrates. X-ray diffraction studies reveal that the films are high-quality, single-phase, and epitaxial (Figure 1a). A zoom-in about the 002-diffraction shows the Ba(5$B$)O and SRO peaks with 002 STO (inset of Figure 1a). The full width at half maximum (FWHM) of the 002 rocking curve is 0.07 for the 120 nm Ba(5*B*)O film shown in Figure 1b, which indicates outstanding crystalline uniformity of the film. Reciprocal space mapping (RSM) studies around the asymmetric (204) Bragg's reflection reveal that the underlying SRO layer is coherently strained to the substrate and that the Ba(5*B*)O film is relaxed due to the high lattice mismatch of ~6% between the film ( a = 4.122 Å) and STO ( a = 3.905 Å).[46]

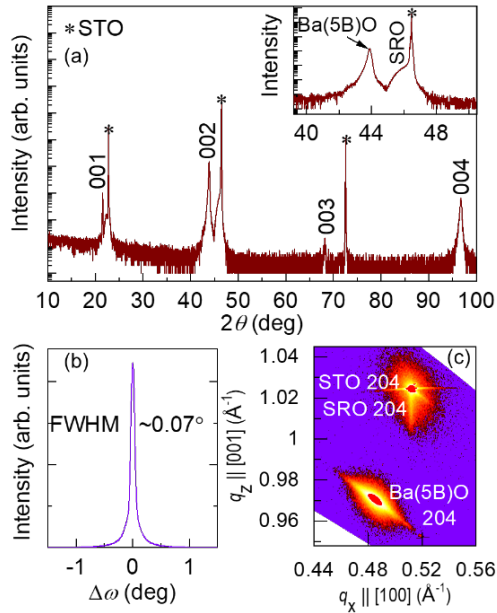

**Figure 1.** (a) $\theta$–$2\theta$ X-ray diffraction line scans of epitaxial Ba(5*B*)O thin films on SRO layered (001) STO substrate. Inset shows a zoom-in about the Ba(5*B*)O (002)-diffraction peak. (b) A rocking curve about the Ba(5*B*)O 002 peak ($\omega$ = 21.944). (c) RSM around the (204) asymmetric reflections of STO, SRO, and Ba(5*B*)O film.

To study the relaxor behavior and the impact of the cation disorder on the dielectric properties of Ba(5*B*)O films, we probed the evolution of the dielectric response as a function of temperature. Temperature dependent dielectric permittivity and loss tangent plots at different ac frequencies are presented in Figures 2a and 2b. The temperature dependent dielectric data are replotted in Figures 2c–2e. Similar to bulk BaTiO$_3$, the films exhibit three distinct phase transitions[61]. These transitions for the Ba(5*B*)O films are attributed to rhombohedral (R)→orthorhombic (O) at ~170 K, O→tetragonal (T) at ~332 K, and T→cubic (C) at ~570 K. The temperature dependent Raman spectroscopic studies on the films further corroborate these three phase transitions, where the observation of unique Raman bands[62] corresponding to their respective four phases evolved with rising temperature (Figure S1, supporting information). The O→T and T→C phase transition are found to be shifted towards a higher temperature compared to the bulk single crystal BTO.[61] Dielectric losses are low until temperatures begin to approach the ferroelectric to paraelectric phase transition at 500 K where losses exponentially increase (Figure 2b). Like dielectric permittivity, the anomalies in the loss tangent can also be observed close to the phase transition temperatures.

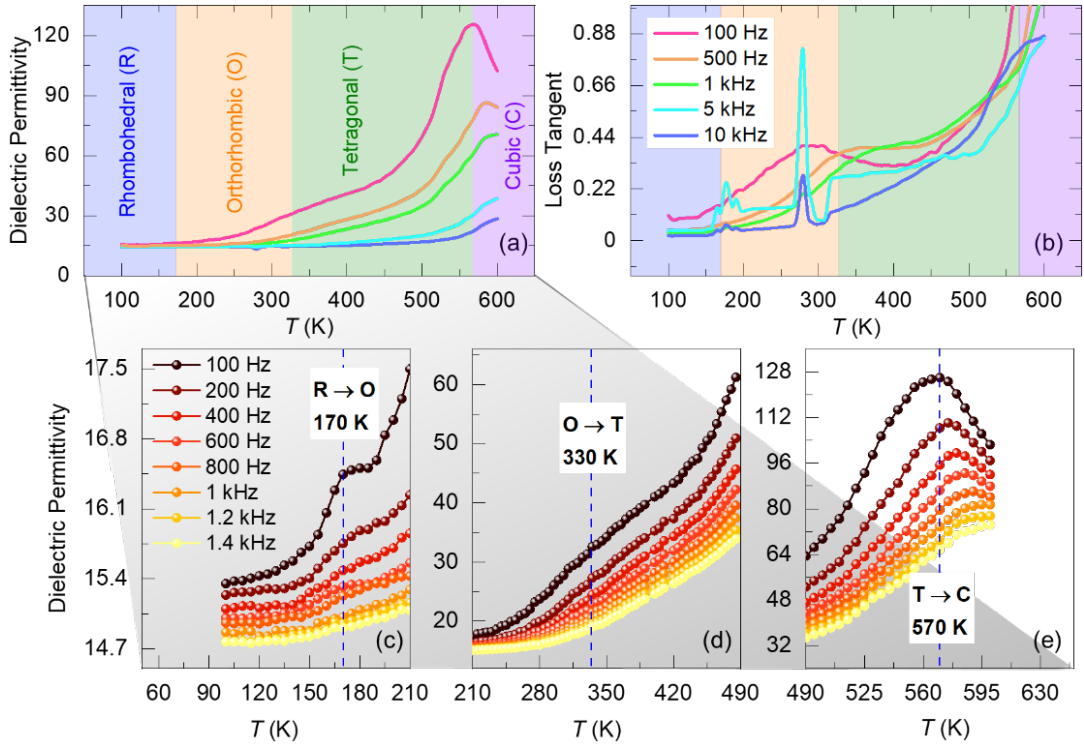

**Figure 2.** Temperature variation of (a) dielectric permittivity and (b) loss tangent at various frequencies (100 Hz–10 kHz) for the Ba(5B)O films. (c–e)Three different phase transitions are clearly observed by zoom-in the temperature variation of dielectric permittivity from (a).

From Figure 2e, the T→C phase transition, which is also correspond to a ferroelectric to paraelectric phase transition with Curie temperature $T_C$ ~570 K, proceeds gradually with temperature rather than sharply. Such diffuse characteristic is a feature of typical relaxor behavior.[8]

The dielectric maximum for the lowest frequency (100 Hz) occurs at the temperature, $T_m \approx 570$ K. As expected for relaxor ferroelectrics, a strong frequency-dependent change in $T_m$, $\Delta T_{disp} = T_m^{100\,Hz} - T_m^{1\,kHz} \approx 30$ K, and strong frequency dispersion below $T_m$ can be observed in the Ba(5B)O films. The relaxor feature of the Ba(5B)O films is described by the diffuseness factor γ, which varies between 1 (for normal ferroelectrics) to 2 (for ideal relaxor ferroelectrics).[63] However, for relaxors, it is also common to find 1 < γ < 2 depending on the strength of the relaxor character.[64] The factor γ can be derived from the modified Curie-Weiss law;

$$\frac{1}{\varepsilon} - \frac{1}{\varepsilon_m} = \frac{(T-T_m)^\gamma}{C}, \tag{1}$$

where $\varepsilon_m$ corresponds to the maximum dielectric permittivity and $C$ is a constant.[63] Figure 3a shows the plot of the modified Curie-Weiss law, where the slope of the plot γ = 1.62 indicative of strong relaxor character of the Ba(5B)O film. The inverse of the dielectric permittivity versus temperature plot is used to obtain the Burns temperature ($T_B$) which is identified as the temperature at which the first deviation from Curie–Weiss behavior appears in the temperature dependent measurement of properties (Figure 3b). $T_B$ is related to the polarization fluctuations due to the formation of polar nanoregions in relaxor ferroelectrics.[14] Unlike normal dielectrics, the frequency dependence of $T_m$ in relaxors deviates from the Arrhenius dependence due to the presence of dipolar correlations[65]. In relaxors, this dependence obeys an empirical Vogel- Fulcher relation;

$$f = f_0 \exp\left(\frac{-E_a}{K_B(T_m-T_f)}\right), \tag{2}$$

where $f_0$ is the limiting response frequency of the dipoles, $f$ is the applied ac field frequency, $E_a$ is the activation energy, $T_f$ is the static freezing temperature, and $K_B$ is the Boltzmann's constant.[65] As shown in Figure 3c, a Vogel–Fulcher relationship with an extrapolated freezing temperature of around 565 K corroborates the relaxor ferroelectric behavior of Ba(5B)O films.[64,66] The values of $E_a$ (~0.16 eV) and $f_0$ (~3.18x10$^{12}$ Hz) extracted from the fit are similar to those values obtained for Pb-[64] and BTO-based relaxors.[39]

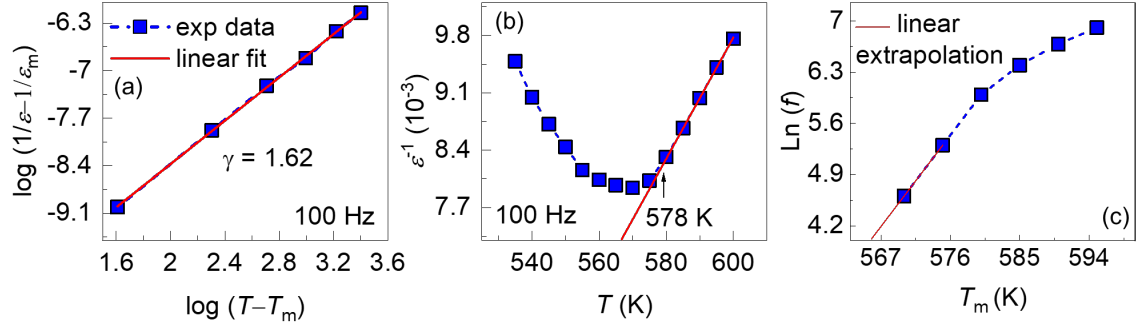

**Figure 3.** (a) Plot of modified Curie-Weiss law and fitting of the relaxor factor γ of Ba(5B)O films. (b) Temperature dependence of the reciprocal of the dielectric permittivity where the straight line reveals that the dielectric response starts to follow the Curie-Weiss law at temperatures higher than $T_m$. Vogel–Fulcher plot with extrapolated freezing temperature ($T_f$ = 565 K, red line), indicating relaxor ferroelectricity in epitaxial Ba(5B)O films.

To gain a deeper understanding of the evolution of polar symmetry through the phase transitions, temperature-dependent optical second harmonic generation (SHG) was used to monitor changes to the crystal's inversion symmetry. Figure 4a shows a schematic diagram of the experimental setup for the SHG measurements in reflection. Figures 4b–d show the s-polarized SHG signal at different temperatures in T, O, and R phases as a function of input fundamental polarization. Theoretically, this s-wave pattern should change depending on crystal symmetry, given by the following equations as functions of the incidence optical polarization angle $\phi$:

$$I_{S-T}(\phi) \sim I_{S-O}(\phi) \sim \sin^2 2\phi, \tag{3}$$

$$I_{S-R}(\phi) \sim A_1(\cos^2\phi + A_2 \sin^2\phi)^2 + A_3 \sin^2 2\phi + A_4 \sin 2\phi\,(\cos^2\phi + A_5 \sin^2\phi), \tag{4}$$

where $A_i$ are the factors determined by tensor components of the second order susceptibility $\chi$, which satisfies $P_i(2\omega) = \varepsilon_0\,\chi_{ijk}\,E_j(\omega)E_k(\omega)$.[67]

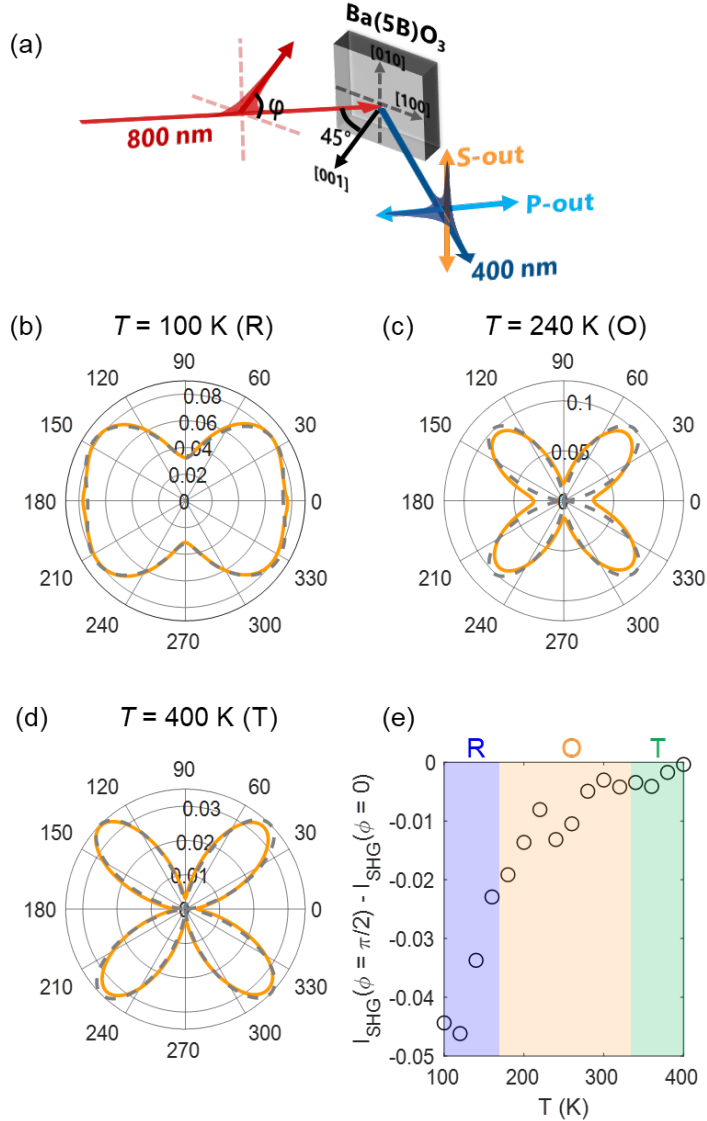

**Figure 4.** (a) Experimental geometry for the SHG measurements. The s-wave reflected SHG signals measured at temperatures of (b) 100 K, (c) 240 K and (d) 400 K. (e) Temperature dependence of the difference in SHG signals between $\phi = 0°$ and $90°$.

Using equations 3 and 4, we fit our experimental data in Figures 4b–d with the grey dotted lines (see supporting information for SHG analysis). The overall agreement of the fitting to the SHG data supports our assignment of crystal symmetries at different temperatures. Particularly, an abrupt change arises from O to R phases, as the clear four-fold symmetry disappears with decreasing temperature (Figure 4b). For a more quantitative analysis, we extract the difference in SHG signals between $\phi = 0°$ and $90°$, which should be zero with a four-fold symmetric pattern of sin (2$\phi$) in T and R phases. From Figure 4e, an anomaly below ~ 170 K in the SHG difference indicates the onset of O→R phase transition, which is in agreement with the temperature dependent dielectric measurements in Figure 2c.

While SHG demonstrates changes in symmetry that are consistent with polar distortions that may signal a ferroelectric character in the Ba(5*B*)O films, room temperature polarization switching and hysteresis loops are better suited to monitoring macroscopic functionality. The polarization-voltage (P-V) and switching current-voltage (I-V) hysteresis loops are measured at a frequency of 100 Hz. Figure 5a shows the recorded hysteresis loops at 300 K; a double hysteresis loop is obtained which indicates the antiferroelectric (AFE) behavior in the sample.[68,69] As can also be seen in Figure 5a, the I-V loop shows four peaks corresponding to the double hysteresis P-V loop, confirming that the AFE behavior is present in the Ba(5*B*)O films.[68]

The atomic-scale microstructures are characterized using high-angle annular darkfield (HAADF) Z-contrast STEM to determine the domain structures according to the projected displacement of *B*-site cation relative to the lattice center of its four nearest neighboring Ba (*A*-site). Figure 5b shows the HAADF-STEM image where the domains with yellow dashed lines can be delineated using the projected *B*-site cation displacements. The domains of projected *B*-site cation displacements in different directions with the size of around 2 to 6 nm can be observed (Figures 5b–e). Similar domain structures of the size of 2-5 nm have been observed in the thin films of BTO-STO-BiFeO$_3$ solid solutions,[16] where the chemical disorder induced by changing the composition of the end-members in the mixture were observed to be the main reason for the formation of nanodomains. It is to be noted that these domains are not pockets of like elements in Ba(5*B*)O films.[46] These results indicate that the *B*-site cation disorder in Ba(5*B*)O samples leads to the observed nanodomains where enhanced local-chemical disorder intensifies the polarization disorder. Follow on local electromechanical properties of epitaxial Ba(5*B*)O films across the phase transition temperatures could be interesting to examine the piezoelectric responses in these highly disorder oxides.

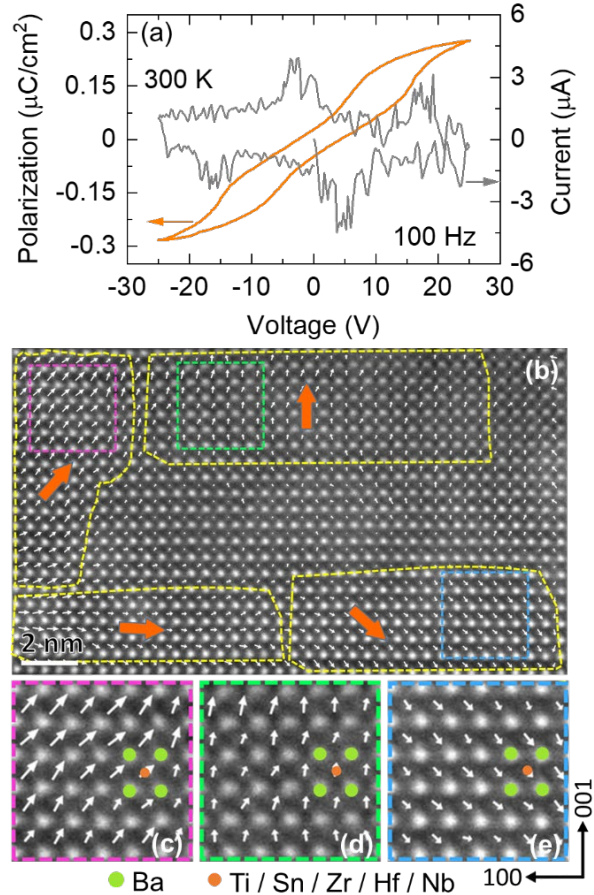

**Figure 5.** (a) The polarization-voltage hysteresis loop measured at a frequency of 100 Hz at 300 K along with the corresponding switching current-voltage characteristic curve for Ba(5*B*)O films. (b) HAADF STEM image of Ba(5*B*)O. The yellow dashed lines delineate the nanodomains, with the projected *B*-site cation displacements denoted by light-blue arrows. (c–e) Magnified images of selected areas from (b) to show the cation displacements. The white arrows show the *B*-site cation displacement vectors in each unit cell.

Density functional theory (DFT) calculations are used to estimate the relative feasibility of formation of other single phase high-entropy perovskite relaxor oxides with different combinations of cations. We have utilized pairwise mixing enthalpies [70,47] to predict the phase stability and cation disorder of Ba(5*B*)O, including other possible combinations of high entropy perovskite oxides (HEPOs) Ba$BO_3$ with $B =$ {Hf, Nb, Sc, Sn, Ta, Ti, Zr}. We have considered the stability of two component perovskites in seven different perovskite phases, 40 atom super cell of the antiferroelectric structure, and non-polar rhombohedral structures with octahedral rotations (Figure S2 and Table S1, supporting information). The pairwise mixing enthalpies $\Delta H_{\text{mix}}[\text{Ba}(BB')O_x]$ in a generic two component perovskite structure, Ba$(BB')O_x$ is estimated using the following equation:

$$\Delta H_{\text{mix}}[\text{Ba}(BB')O_x] = E_{\text{DFT}}[\text{Ba}(BB')O_x] - E_{\text{DFT}}[\text{BaO}, G]$$

$$- \frac{1}{2p} E_{\text{DFT}}[B_p O_y, G] - \frac{1}{2q} E_{\text{DFT}}[B'_q O_z, G] - \frac{1}{2}\left(x - \frac{y}{2p} - \frac{z}{2q}\right) E_{\text{DFT}}[O_2, G], \tag{5}$$

where $E_{\text{DFT}}[\text{Ba}(BB')O_x]$, $E_{\text{DFT}}[\text{BaO}, G]$, $E_{\text{DFT}}[B_p O_y, G]$, $E_{\text{DFT}}[B'_q O_z, G]$, $E_{\text{DFT}}[O_2, G]$ are the total energies of the two component perovskite, BaO, $B_p O_y$, $B'_q O_z$ and $O_2$ in their ground phase structures. Figure 6(a) plots the values of $\Delta H_{\text{mix}}[\text{Ba}(BB')O_x]$, in the non-polar cubic $Pm\bar{3}m$ phase. Each compound is annotated with 'C', 'R', 'T' or 'A', representing the ground state cubic, rhombohedral, tetragonal and antiferroelectric phases, respectively. Similar plots for mixing enthalpies in the 6 other phases considered in this study are presented in the Supporting Information (Figure S3). We find that the cubic structure is the ground state phase for 6 one-component and 18 two-component combinations out of 7 (diagonal) and 21 (off-diagonal) compounds, respectively. For BTO, the ground state is rhombohedral, agreeing with previous theoretical[71] and experimental studies.[72] Furthermore, the polar tetragonal phase was found to be the ground state for the following two component perovskite oxides – Ba(HfNb)$O_3$ and Ba(NbSn)$O_3$.

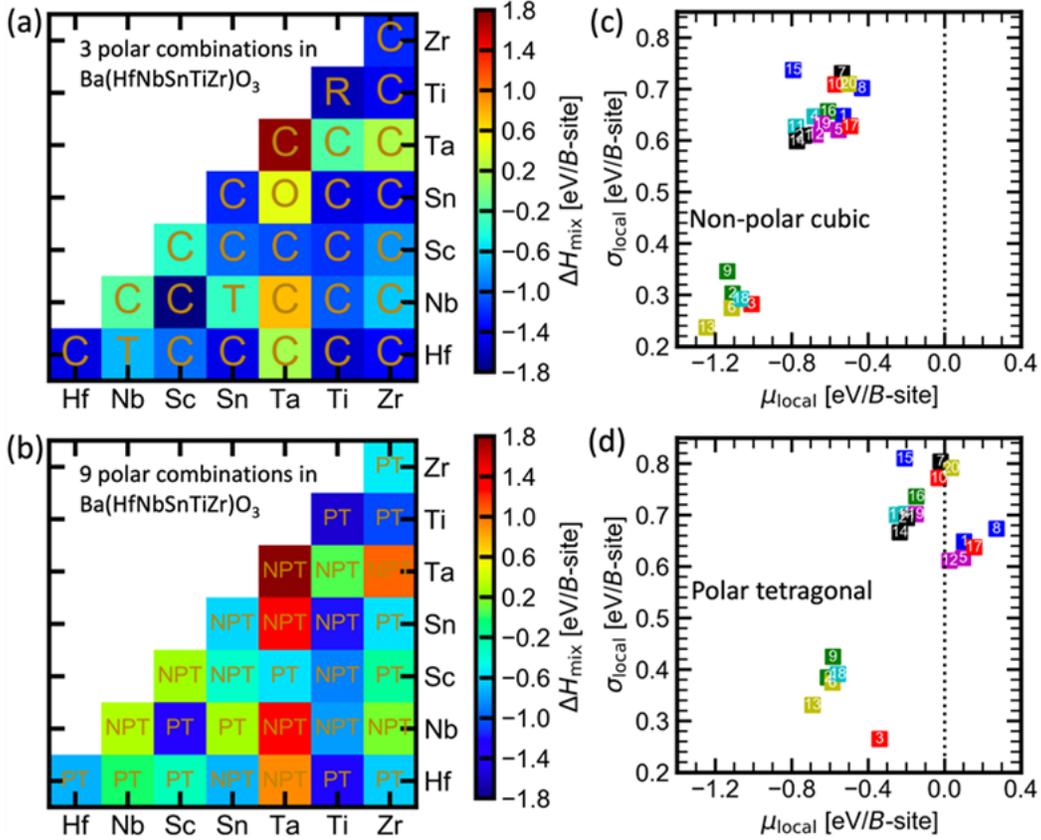

**Figure 6.** (a) Mixing enthalpies, $\Delta H_{mix}$ of $BaBO_3$ perovskite oxides (diagonal) and two component perovskite oxides, $Ba(BB')O_3$ in the relaxed non-polar cubic phase. $B$-sites chosen from a set of $B = \{$Hf, Nb, Sc, Sn, Ta, Ti, Zr$\}$. The annotations 'C', 'T', 'R' and 'A' represent the ground state cubic, tetragonal, rhombohedral, and antiferroelectric phases of each compound, respectively. (b) The values of $\Delta H_{mix}$ for the epitaxially constrained non-polar tetragonal (NPT) and polar tetragonal (PT) phases. (c, d) Comparison of the enthalpy and entropy descriptors, $\mu_{local}$ and $\sigma_{local}$, respectively for all 21 five component combinations in relaxed and epitaxially constrained structures. The five-component perovskite represented by 2, 3, 6, 9, 13 and 18 doesn't contain $Ta^{5+}$ and could be potentially synthesized as high entropy perovskites in both relaxed and epitaxially constrained coherent thin films.

We utilize enthalpy and entropy descriptors to compare the relative formation abilities of 21 five component perovskites. These thermodynamic descriptors are estimated by systematically exploring all possible local configurations. The mean and standard deviation of the local mixing enthalpies are the enthalpy and entropy descriptors, $\mu_{local}$ and $\sigma_{local}$, respectively.[70] Essentially, the descriptors for the enthalpy and entropy, are obtained from exploring the local mixing enthalpies, of all local configurations within the five component perovskites. When the distribution of mixing enthalpies of local configurations is relatively narrow, indicating a small value for $\sigma_{local}$, consequently, the entropy contribution ($-T\Delta S$) to the free energy of the high entropy structure (which is inversely proportional to the entropy descriptor, $\sigma_{local}$) would be large and negative. Furthermore, lower values for the enthalpy descriptor, further stabilize the free energy. Hence, a five-component oxide (FCO) with relatively low values for $\sigma_{local}$ and $\mu_{local}$, should have a relatively high propensity for the formation of a phase pure high entropy structure. Figure 6(b) compares the values of $\mu_{local}$ and $\sigma_{local}$ for all 21 five component combinations annotated by their index. Table 1 presents the indices and corresponding five component perovskites. The five-component perovskites 2, 3, 6, 9, 13 and 18 that do not contain $Ta^{5+}$, forming a cluster at the bottom left corner, have relatively low $\mu_{local}$ and $\sigma_{local}$ and could be potentially synthesized as high entropy perovskites. Our method accurately predicts that the FCO 9, discussed in the experimental section of this part, $Ba(HfNbSnTiZr)O_3$, would form as a high entropy perovskite. In addition, we also predict the following five component combinations to have relatively high potential to form as high entropy perovskites—2, 3, 6, 13 and 18—obtained by replacing one of the $B$-site element Zr, Ti, Sn, Nb or Hf by Sc. The second cluster, at the top right, containing $Ta^{5+}$ has a relatively large $\sigma_{local}$. Due to the large value of the entropy descriptor, these five component combinations are unlikely to form as a high entropy perovskite. The enthalpy and entropy descriptors plots for the 6 other phases considered in this study are presented in the Supporting Information (Figure S4).

**Table 1.** The indices and corresponding five component perovskites.

| Index | Perovskite | Index | Perovskite | Index | Perovskite |
|---|---|---|---|---|---|
| 1 | $Ba(HfNbScSnTa)O_3$ | 8 | $Ba(HfNbSnTaZr)O_3$ | 15 | $Ba(HfSnTaTiZr)O_3$ |
| 2 | $Ba(HfNbScSnTi)O_3$ | 9 | $Ba(HfNbSnTiZr)O_3$ | 16 | $Ba(NbScSnTaTi)O_3$ |
| 3 | $Ba(HfNbScSnZr)O_3$ | 10 | $Ba(HfNbTaTiZr)O_3$ | 17 | $Ba(NbScSnTaZr)O_3$ |
| 4 | Ba(HfNbScTaTi)O$_3$ | 11 | Ba(HfScSnTaTi)O$_3$ | 18 | Ba(NbScSnTiZr)O$_3$ |
| 5 | Ba(HfNbScTaZr)O$_3$ | 12 | Ba(HfScSnTaZr)O$_3$ | 19 | Ba(NbScTaTiZr)O$_3$ |
| 6 | Ba(HfNbScTiZr)O$_3$ | 13 | Ba(HfScSnTiZr)O$_3$ | 20 | Ba(NbSnTaTiZr)O$_3$ |
| 7 | Ba(HfNbSnTaTi)O$_3$ | 14 | Ba(HfScTaTiZr)O$_3$ | 21 | Ba(ScSnTaTiZr)O$_3$ |

We have also evaluated the influence of epitaxial strain on the stability of the high entropy phase as well as its polar properties. It is crucial to investigate the formation ability and polar properties of a 'coherent' thin film – where the in-plane lattice constant is fixed to the lattice constant of the STO substrate – because the epitaxial strain is known to influence the functional properties and stability in perovskites.[73,74] The DFT optimized lattice constant of STO and BTO in the cubic phase are 3.859 Å and 3.961 Å, respectively, suggesting that the coherent thin film would be in epitaxial compression $\varepsilon \cong -2.6\%$. We find that for phases with in-plane polar displacements—i.e., rhombohedral $R3m$, orthorhombic $Amm2$ and monoclinic $Cm$—when relaxed in DFT with fixed epitaxial strain, result in vanishing in-plane polar displacements. This observation is in agreement with the vanishing in-plane polarization components in perovskites under epitaxial compression.[75] Due to the vanishing in-plane polar components, we find that there are only two stable epitaxial phases – non-polar tetragonal (NPT) $P4/mmm$ and polar tetragonal (PT) $P4mm$ phases. Figure 6(c) presents the values of $\Delta H_{\text{mix}}[\text{Ba}(BB')O_x]$, in NPT and PT phases. The non-polar phase is tetragonal $P4/mmm$ instead of cubic $Pm\bar{3}m$ because of the symmetry breaking of the three-fold rotation axis imposed by epitaxial constraint. The mixing enthalpy values of the epitaxial structures are generally greater than the values of the relaxed structures. Another crucial difference between the relaxed and epitaxial structures is the number of two-component polar combinations (TCPCs)—only 3 TCPCs (Ti-Ti, Hf-Nb and Nb-Sn) are stable in the case of relaxed structures, whereas 9 TCPCs (Ti-Ti, Hf-Hf, Zr-Zr, Hf-Nb, Hf-Ti, Hf-Zr, Nb-Sn, Sn-Zr and Ti-Zr) are stable under epitaxial compression. The epitaxial strain stabilizes the polar structures for 60% of TCPCs—a striking 40% increase from 20% (3 TCPCs). Due to the increase in TCPCs, more microstates could be polar in the case of a coherent epitaxially compressed film than the relaxed thin film, leading to improved polar properties in the HEPO. Figure 6(d) compares the values of $\mu_{local}$ and $\sigma_{local}$ for all 21 five component combinations under epitaxial compression, annotated by their index. Similar to the relaxed structures, we find that the five-component perovskites 2, 3, 6, 9, 13 and 18 that do not contain $Ta^{5+}$, forming a cluster in the bottom left corner, have relatively low $\mu_{local}$ and $\sigma_{local}$. Even though the values of the thermodynamic descriptors are larger for the epitaxially-strained, coherent thin films than their relaxed counter parts, we believe that these five component combinations can still be synthesized as high entropy relaxor ferroelectric thin films.

## 4. CONCLUSION

We characterized the dielectric and ferroelectric properties of epitaxial $Ba(Ti_{0.2}Sn_{0.2}Zr_{0.2}Hf_{0.2}Nb_{0.2})O_3$ films across the phase transition temperatures where the strong limit of compositional complexity within the lattice has a dominating role in crystal phase metastability and Curie temperature. The observation of polar nano-domains and diffuseness of the phase transition at Curie temperature confirms the relaxor ferroelectric nature of Ba(5*B*)O films. Our study implies that the deterministic creation of local configurational disorder could be an effective way to enhance functional responses in perovskite oxides and provide an approach to the design of new relaxor ferroelectrics. Furthermore, computational prediction of formation feasibility of five more high entropy perovskites in relaxed and epitaxial compression, with Ba on *A*-site, suggests that the rich chemical space is yet to be explored for their functionalities.

## ASSOCIATED CONTENT

**Supporting information:** Temperature dependent Raman measurements, analysis of SHG patterns, crystal structures considered in the DFT work and mixing enthalpies for those structures, and calculations for comparison of the difference of enthalpy and entropy descriptors.

## AUTHOR INFORMATION

**Corresponding author**

Yogesh Sharma (ysharma@lanl.gov)

**Notes:** The authors declare no competing financial interest.

## ACKNOWLEDGMENTS

This work was supported by the U.S. Department of Energy (DOE), Office of Science, Basic Energy Sciences (BES), Materials Sciences and Engineering Division (conception, synthesis, structural characterization, and theory). Optical characterization performed at Los Alamos National Laboratory was supported by the G. T. Seaborg Institute under project number 20210527CR and the NNSA's Laboratory Directed Research and Development Program and was performed, in part, at the Center for Integrated Nanotechnologies, an Office of Science User Facility operated for the U.S. Department of Energy Office of Science. Los Alamos National Laboratory, an affirmative action equal opportunity employer, is managed by Triad National Security, LLC for the U.S. Department of Energy's NNSA, under contract 89233218CNA000001. Some microscopy measurements were conducted through user proposal at the Center for Nanophase Materials Sciences, which is a US DOE, Office of Science User Facility. KKM and RSK acknowledge financial support from the Department of Defense, USA (DoD Grant #FA9550-20-1-0064). This manuscript has been authored by UT-Battelle, LLC under Contract No. DE-AC05-00OR22725 with the U.S. Department of Energy. The United States Government retains and the publisher, by accepting the article for publication, acknowledges that the United States Government retains a non-exclusive, paid-up, irrevocable, world-wide license to publish or reproduce the published form of this manuscript, or allow others to do so, for United States Government purposes. The Department of Energy will providepublic access to these results of federally sponsored research in accordance with the DOE Public Access Plan (http://energy.gov/downloads/doe-public-access-plan).

## AUTHORS CONTRIBUTIONS

The manuscript was written through contributions of all the authors. All authors have read and agreed to the final version of the manuscript.